\documentclass[onecolumn, showpacs, showkeys]{revtex4}
\usepackage{graphicx}
\usepackage{amsmath, amsthm, amsfonts,amssymb}
\usepackage{graphicx}

\begin{document}
\title{Deterministic Factors of Stock Networks
based on Cross-correlation in Financial Market}

\author{Cheoljun \surname{Eom$^{1}$}} \thanks{Electronic address: shunter@pusan.ac.kr}
\author{Gabjin \surname{Oh$^{2}$}} \thanks{Electronic address: gq478051@postech.ac.kr}
\author{Seunghwan \surname{Kim$^{2}$}} \thanks{Electronic address: swan@postech.ac.kr}

\affiliation{$^{1}$Division of Business Administration, Pusan
National University, Busan 609-735, Korea} \affiliation{$^{2}$Asia
Pacific Center for Theoretical Physics \& NCSL, Department of
Physics, Pohang University of Science and Technology, Pohang,
Gyeongbuk, 790-784, Korea}

\received{10 01 2007}

\begin{abstract}

The stock market has been known to form homogeneous stock groups
with a higher correlation among different stocks according to common
economic factors that influence individual stocks. We investigate
the role of common economic factors in the market in the formation
of stock networks, using the arbitrage pricing model reflecting
essential properties of common economic factors. We find that the
degree of consistency between real and model stock networks
increases as additional common economic factors are incorporated
into our model. Furthermore, we find that individual stocks with a
large number of links to other stocks in a network are more highly
correlated with common economic factors than those with a small
number of links. This suggests that common economic factors in the
stock market can be understood in terms of deterministic factors.

\pacs {89.65.Gh, 89.75.Fb, 89.75.Hc} \keywords {stock network,
minimal spanning tree, multi-factor model, econophysics}
\end{abstract}

\maketitle

\section{Introduction}

Stock markets have long been known to be extremely complex systems,
evolving through interactions between heterogeneous units. Hence,
the attempts to study and  understand the nature of interactions
between stocks have been important in understanding the pricing
mechanism in the stock market. The cross-correlation matrix has been
widely used to quantify interaction among stocks. If we can classify
and use significant information included in the cross-correlation
matrix between stocks, we will better understand the stock market.
However, the extraction of significant information from the
cross-correlation matrix has been quite difficult. In finance,
researchers have usually used the methods of multivariate
statistical analysis, such as principal component analysis, factor
analysis, and cluster analysis. In econophysics, for instance, a
stock network is proposed by Mantegna {\em et al.} for investigating
the interaction between stocks using the minimal spanning tree (MST)
method [1].

The stock network visually constructs the relationship between
stocks, which is extracted by the MST based on the
cross-correlations between stock returns. Our work is based on the
arbitrage pricing model (APM), widely acknowledged in financial
literature [2]. That is, there are many common economic factors in
the stock market, which influence all stocks traded [3]. Common
economic factors include the industrial product, the risk premium,
the term structure of interest, and inflation. The stocks with the
same common economic factors are highly correlated with each other
and tend to be grouped into a community. That is, individual stocks
are divided into small homogeneous stock groups depending on the
tendency of stock price changes in correlation with common economic
factors [4-6]. Accordingly, the pricing mechanism of individual
stocks might be explained by the common economic factors in the
stock markets. These observations in the financial sector are
similar to the results derived from the MST that in a stock network
individual stocks belonging to the same industry from groups with
linking relations [7-10].

In previous studies, much focus is made on the topological
properties and the formation principles of stock networks, which
revealed that the degree distribution of a stock network follows a
power law [11-12]. This implies that most individual stocks in a
network have a small number of links with other stocks, while a few
stocks, so called hub stocks, have a large number of links. Eom {\em
et al.} suggested that the larger the degree of a stock is in the
stock network, the more it is correlated with the market index
empirically [13]. Therefore, some stocks acting as a hub for each
cluster in a network are affected much more by the market index.

In a recent work, Bonnano {\em et al.} investigated the degree of
consistency between networks using estimated returns from the
pricing model and networks from real stock returns [14]. The purpose
of this study is to investigate whether stock returns from pricing
models can explain interactions between stocks. This study may
reveal the deterministic factors that significantly affect the
formation process of a stock network. The widely accepted model in
the previous study has been the capital asset pricing model (CAPM)
[15-16]. In the CAPM, the prices of individual assets are determined
by the market portfolio including all risk assets. In addition, the
one-factor model (or the market model), which is a empirical model
of the CAPM, uses a market index as a proxy for the market
portfolio. However, stock networks from the estimated returns of the
one-factor model show a very different structure from the original
stock network [9-10, 14]. That is, even if the market index as a
representative factor is important for determining the stock price,
the market index alone cannot completely explain interactions among
stocks.

Therefore, the interactions between stocks may be explained better
by the multi-factor model than the one-factor model. As the APM
proposed by Ross suggests, stock prices are mostly determined by
common economic factors observed in stock markets. These results
show that the grouping process of stocks can be observed
systematically by the cross-correlation matrix between stocks [17].
The previous study also found that an estimated stock network with
returns generated by the stochastic dynamics model (with control
parameters included in the market, group and individual stock
properties) is very similar to the original stock network with real
return data of the stock market [18].

In order to find deterministic factors of the stock network, we
propose a model based on common economic factors and investigate
whether these common economic factors in the stock market play an
important role in determining the stock network. We used the APM,
extensively acknowledged as the multi-factor model in financial
literature, so that the estimated stock returns reflect the
properties of common economic factors. In addition, we investigate
the degree of consistency between the original stock networks with
real return data and the estimated network with returns based on the
multi-factor model. To quantify the strength of consistency between
stock networks, we use the survivor ratios suggested in the previous
studies [19]. This ratio measures whether stocks that are linked
directly to specific stocks in the original stock network have the
same links to those in the estimated stock network. We found that
the estimated stock networks with more common economic factors in
the stock market, give a higher consistency with the original stock
network of real returns. In particular, in stock networks, stocks
with a large number of links to other stocks in the original stock
network have a higher consistency than those with a small number of
links. Therefore, these results suggest that common economic factors
in the stock market may help to explain the formation principles of
stock networks.

In the next section, we describe the data and methods used in this
paper. In Section III, we present the results obtained from the APM.
Finally, we end with a summary and conclusion.

\section{Data and methods}

We used the daily prices of $N=400$ individual stocks in the S\&P
500 index of the American stock market from January 1993 to May
2005. The test procedure in this paper can be explained by the
following three steps. First, we determine input data that is needed
to create the stock network. Second, we create a stock network using
the MST method. Third, we calculate the survivor ratio to compare
stock networks.

In the first step, the return from input data is significant in that
we use real stock returns $R_{j}=ln(P_{t+1})- ln(P_{t})$ and
estimated stock returns $\widehat{R}$ by using the multi-factor
model, respectively. The correlation matrix $\rho_{ij}$ using real
and estimated returns is calculated by

\begin{equation}\label{e1}
 \rho_{ij} \equiv \frac{\langle R_{i}R_{j} \rangle - \langle R_{i}\rangle \langle R_{j}\rangle}{\sqrt{({\langle R_{i}^{2}\rangle}-{\langle R_{i}\rangle}^{2})
 ({\langle R_{j}^{2}\rangle}-{\langle R_{j}\rangle}^{2})    }},
\end{equation}
where the notation $\langle \cdots \rangle$ means an average over
time. The cross-correlation coefficient can lie in the range of $-1
\leq \rho_{ij} \leq +1 $, where $\rho_{ij}=-1$ denotes completely
anti-correlated stocks and $\rho_{ij}=1$, completely correlated
stocks. In the case of $\rho_{ij}=0$, the stocks $i$ and $j$ are
uncorrelated.

In the second step, we create a stock network with a significant
relationship between stocks using the MST method. The MST, a
theoretical concept in graph theory [20], is the spanning tree of
the shortest length using the Kruskal or Prim algorithm [21-22].
Therefore, it is a graph without a cycle connecting all nodes with
links. This method is also known as the single linkage method of
cluster analysis in multivariate statistics [23]. The metric
distance $d_{ij}$, introduced by Mantegna, relates the distance
between two stocks to their cross-correlation coefficient [24], and
is defined as

\begin{equation}\label{e2}
d_{ij} \equiv \sqrt{2(1-\rho_{ij})},
\end{equation}
where the distance $d_{ij}$ can lie in $ 0 \leq d_{ij} \leq 2$,
where the small distance implies strong cross-correlation between
stocks. In addition, this distance matrix can be used to construct
stock networks using essential information of the market.

In the third step, we measure the degree of consistency between
stock networks generated by models based on common economic factors
and those for the original stock markets. The stock networks studied
are the original stock network $G^{o}$, created using the reals
return data and the estimated stock networks $G^{E}$, using returns
from the multi-factor model. To quantify the consistency between
networks, we use the survivor ratio. This is a ratio of frequency
$FQ$ measuring that stocks that are directly linked with a specific
stock, $j$, in the original stock networks $FQ_{j}[G^{o}_{L}]$ have
the same links with one in the estimated stock network
$FQ_{j}[G^{o}_{L} \bigcap G^{E}_{L}]$ among the number of all
possible connections between stocks in the stock network ($N-1$).
The survivor ratio is defined by

\begin{equation}\label{e3}
 r_{s_{L \geq i}} \equiv \frac{1}{N-1} \sum_{j=1}^{N-1} \frac{FQ_{j}[G^{o}_{L}
\bigcap G^{E}_{L}]}{FQ_{j}[G^{o}_{L}]}, ~(i=1,2,...,M), \\
\end{equation}
where N is the number of stocks with more than the degree i in the
original stock network and the survivor ratio can vary between
$0\leq r_{s} \leq 1$. If $r_{s}=0$, two stock networks have a
completely different structure, and if $r_{s}=1$, they have the same
structure. We calculated the survival ratio according to the various
number of links with other stocks in the original stock network. The
number of links with other stocks, $i$, is from over one $L \geq 1$
to over maximum $L \geq M$. That is, the survivor ratio of $L \geq
1$ targets stocks which have one or more links. Therefore, it
represents a degree of consistency for all stocks which exist in a
given network. The survivor ratio of $L \geq M$ targets stocks that
have the largest number of links; therefore, it shows a degree of
consistency for stocks having the greatest number of links in a
network (, so-called hub stocks).

\section{Results}

\begin{figure}[h]

\includegraphics[height=14cm, width=16cm]{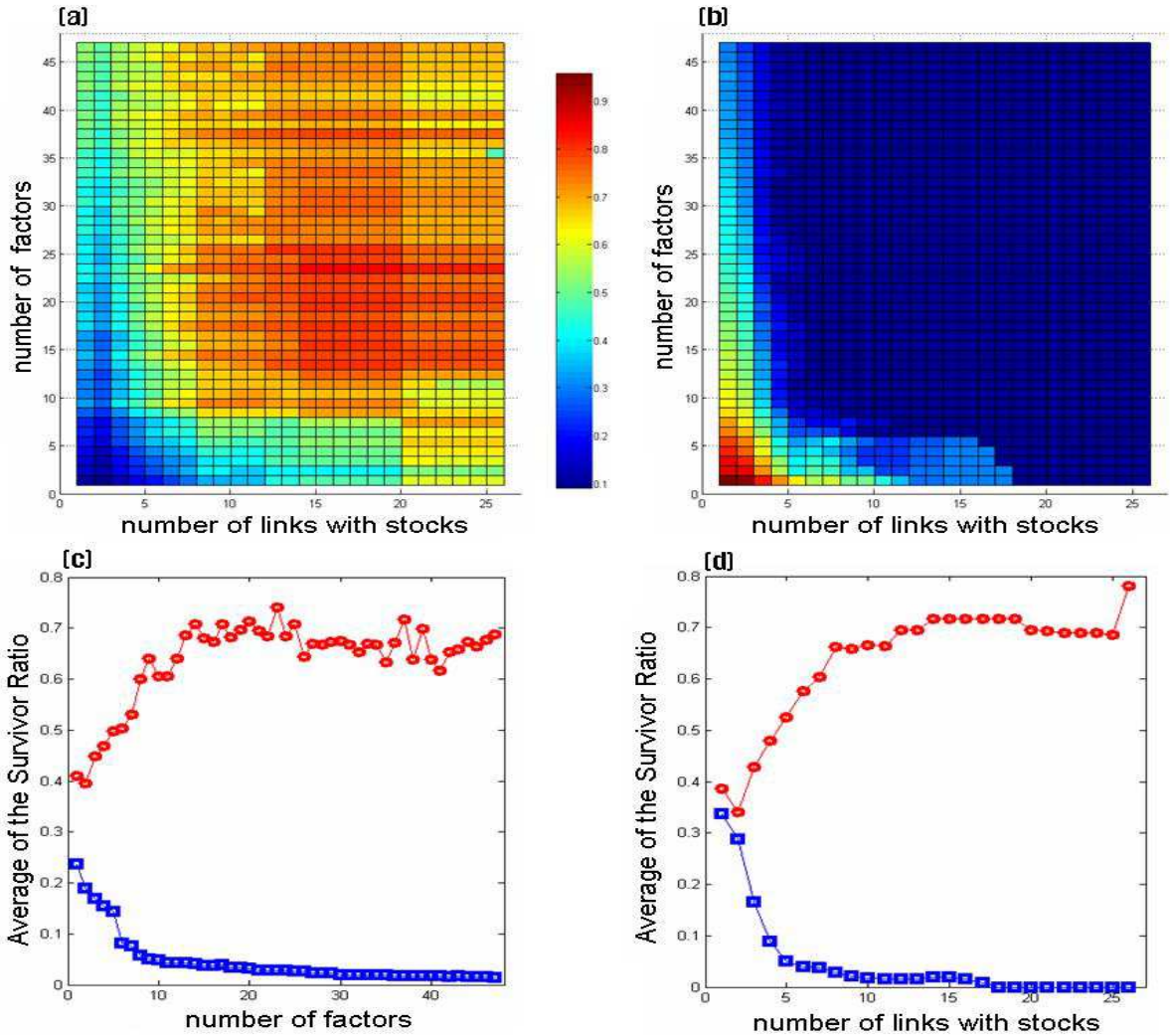}

\caption{\label{fig:consistency} The strength of consistency between
the original stock networks with real returns and estimated stock
networks with (a) estimated returns and (b) residual returns by the
multi-factor model, respectively. Each cell denotes the average of
the survival ratio for the number of common economic factors and the
number of links, respectively. Fig. 1(c) indicates the changes in
the average values of the survivor ratios of all the link numbers
depending on the increase in the number of common economic factors,
irrespective of the link numbers, from the verification results of
Fig. 1(a) of estimated returns and Fig. 1(b) of residual returns.
Fig. 1(d) shows the changes in the average value of the survivor
ratios depending on the increase in the link number, irrespective of
the number of common economic factors, from the verification results
of Fig. 1(a) and Fig. 1(b). In the figure, the circles (red) and
boxes (blue) denote the survival ratio for stock networks with
estimated returns and residual returns, respectively.}
\end{figure}

In this section, using the cross-correlation-based MST method, we
investigated the strength of consistency between the original and
estimated stock networks. The independent variables used in the
multi-factor model are common economic factors that are estimated
through the factor analysis method in multivariate statistics [25].
Factor analysis widely used in the field of social science can
reduce many variables in a given data set to a few factors.

Using factor analysis, we calculated the eigenvalues for returns,
and chose significant factors. We also made new time series called
factor scores in statistics, which use weights with the elements of
an eigenvector. To give economic meaning to significant factors,
regression analysis was conducted with a calculated factor scores
set as dependent variables and the financial market and economic
data as independent variables. After confirming statistically
significant independent variables, factor scores are regarded as
having the attributes of significant independent variables. Through
this process, factor scores can be interpreted as common economic
factors. The common economic factors generated by factor analysis
are used as independent variables in the multi-factor model.
Therefore, the stock returns, $R_{j}(t), ~ j=1,2,...,N$ can be
explained by common economic factors, $F_{k}(t), ~ k=1,2,...,K$, and
the multi-factor model is defined by

\begin{equation}\label{e4}
R_{j}(t) = \alpha_{j} + \beta_{j,1}F_{1}(t) + \beta_{j,2}F_{2}(t) +
...... + \beta_{j,k}F_{k}(t) + \epsilon_{j}(t),
\end{equation}
where $\alpha_{j}$ is an expected return on the stock, $\beta_{j,k}$
are the sensitivity of the stock to changes in common economic
factors, and $\epsilon_{j}(t)$ are the residuals ($E(\epsilon_{j})
\approx 0$, $E(\epsilon_{j}, \epsilon_{m}) \approx 0$, and
$E(\epsilon_{j}, F_{k}) \approx 0$). To establish the multi-factor
model in Eq. 4, we have to determine a number of common economic
factors $K$ and control the problem of multicolinearity between
common economic factors ($Cor(F_{i}(t), F_{j}(t))\approx 0$) [26].

In statistics, there are two methods that determine the number of
significant factors. First, the eigenvalues with the modulus larger
than one determine the significant factors according to the Kaiser
rule [27]. Second, these eigenvalues are arranged in the order of
their size; that is, from the largest value to the smallest one.
After this, the changing slopes between eigenvalues are observed,
and then the point where the changing slopes abruptly become smooth
determine the significant factors. [28]. While the previous work in
the financial field chose the number of common economic factors from
the second method, we will use the first method in order to
investigate results more extensively. Next, when we use the multiple
regression model, there is a problem of multicolinearity among
independent variables. That is, the results may be distorted due to
the higher correlation between independent variables. Therefore, in
order to minimize the correlation between common economic factors,
we created a new time series according to factor analysis controlled
by the rotated varimax method. Here, the maximum number of common
economic factors chosen by the Kaiser rule is $47$. That is, we
consider the number of common economic factors, $F_{k=1}^{k(i)},$
with $i = 1,2,...47$ where $k(i)$ is a number of factors used. To
apply different common economic factors to Eq. 4, we generate new
time series (factor scores) by optimizing through factor analysis
using the method mentioned above. We find that for all cases studied
in our paper the correlations between common economic factors are
very small with the mean correlation of 0.69 \%. That is, there is
no problem with multicolinearity that may occur in multiple
regression models.

We investigate stock networks with returns estimated by the
multi-factor model reflecting various common economic factors. The
stock returns, $\widehat{R_{j}(t)}$, estimated by the multi-factor
model consist of the returns, $\widehat{R^{C}_{j}(t)}$, described by
common economic factors and the returns, $\widehat{R^{R}_{j}(t)}$,
following the random process that cannot be explained by  common
economic factors. The process of creating stock returns can be
explained by the following five steps. First, we determined the
specific number of common economic factors, $F_{k=1}^{k(i)}$.
Second, we estimated the coefficients $\widehat{\alpha_{j}}$ and
$\widehat{\beta_{j,k}}$ of a multi-factor model for all individual
stocks as dependent variables, using common economic factors
($F_{k}$) as independent variables. Third, we calculated the
returns, $\widehat{R^{C}_{j}(t)}=\widehat{\alpha_{j}}+
\sum_{k=1}^{K} \widehat{\beta_{j,k}}F_{k}(t)$, that can be explained
by common economic factors for all individual stocks using the
coefficients estimated in the second step. Fourth, we created the
random returns, $\widehat{R^{R}_{j}(t)}$, with a calculation of the
mean and standard deviation of individual stock returns. Fifth, the
returns, $\widehat{R_{j}(t)} =
\widehat{R^{C}_{j}(t)}+\widehat{R^{R}_{j}(t)}$, can be calculated
from the addition of two returns created in the above process. After
these processes are completed, we created the stock network,
$G^{E}$, with the estimated return of individual stocks by the MST
method. Additionally, we measured the strength of consistency
between the original stock network, $G^{O}$, with real returns and
the estimated stock network, $G^{E}$, with returns created by the
multi-factor model, using the survivor ratio $r_{s}(G^{o},G^{E})$.
The above processes are repeated for all of the common economic
factors from $1$, $F_{k=1}^{k(1)}$, to 47, $F_{k=1}^{k(47)}$,
respectively. Also, as in the above process, we investigated stock
networks with residual returns, with $\epsilon_{j}(t)
=R_{j}(t)-(\widehat{\alpha_{j}} + \sum^{K}_{k=1}
\widehat{\beta_{j,k}}F_{k}(t))$ eliminating the properties of common
economic factors.

Fig. 1 shows the strength of consistency between the original stock
network, $G^{O}$, with real returns and the estimated stock network,
$G^{E}$, with returns created by the multi-factor model using the
measurement of survivor ratio. Fig. 1(a) indicates the survivor
ratios which show the strength of consistency between a stock
network $G^{E}$ derived from estimated returns $\widehat{R_{j}(t)}$
and a stock network $G^{O}$ derived from real returns $R_{j}(t)$. In
the figure, each cell represents the average of the survivor ratios
as the number of common economic factors vary for each number of
links between stocks confirmed in the original stock network. That
is, the $x$-axis denotes the number of links with other stocks $L
\geq i, ~i=1,2,...,49$, in the original stock network. The largest
number of links confirmed in the original stock network was 49
($i=49$) and the second largest number of links was 25 ($i=25$). As
we measured the survivor ratios of stocks with $i$ or more links, we
discovered that the survivor ratio of stocks with $26$ or more links
is the same as that of stock with $49$ or more links. Accordingly,
the links of $x$-axis have been presented from $i=1$~ ($L \geq1$) to
$i=26$ ~($L\geq26$). The $y$-axis denotes the number of common
economic factors, $F_{k=1}^{k(j)}, ~~ j=1,2,...,47$ in the
multi-factor model. In the same way as in Fig. 1(a), Fig. 1(b) shows
the strength of consistency between a stock network derived from
residual returns ($\epsilon_{j}(t)$) and the original stock network.
Fig. 1(c) is used to confirm changes in the average values of the
survivor ratios of the all links depending on the increase in the
number of common economic factors, irrespective of the link number,
from the verification results of Fig. 1(a) using estimated returns
and Fig. 1(b) using residual returns. Meanwhile, Fig. 1(d) is used
to confirm the changes in the average values of the survivor ratios
depending on the increase in link number, irrespective of the number
of common economic factors, from the verification results of Fig.
1(a) and Fig. 1(b). In addition, in Figs. 1(c) and (d), the circles
(red) and boxes (blue) denote the survival ratio for stock networks
with estimated returns, Fig. 1(a), and  residual returns, Fig. 1(b),
respectively.

According to the results, the survivor ratio between the original
stock network and  estimated stock network with estimated returns
had a higher value as the number of common economic factors
increased, while the survivor ratio for stock networks with residual
returns excepting the properties of common economic factors
decreased sharply. These results suggest that if the estimated stock
network can be reflected in the properties of common economic
factors sufficiently, we can observe the interaction between stocks
in the real stock market from the stock network estimated by the
multi-factor model. That is, common economic factors in the stock
market are significantly deterministic factors in terms of making
the stock network. These results were confirmed from Fig. 1(c),
presenting changes in the survivor ratio with estimated returns and
residual returns, respectively, as the number of common economic
factors increase. While estimated returns (the circles)
consecutively reflect the attributes of more common economic factors
in proportion to the increase in the number of common economic
factors, residual return (the squares) exclude the attributes of
common economic factors corresponding to the increasing number of
common economic factors. According to the results, as the number of
common economic factors increased, the survivor ratio depending on
estimated returns also increased but the survivor ratio depending on
residual returns clearly decreased. Therefore, if we use estimated
returns fully reflecting the attributes of common economic factors
from the multi-factor model, we could predict the behavior of the
original stock network. However, we cannot find the rising patterns
of the survivor ratio after the number of common economic factors
exceed certain constant levels ($F^{k(i)}_{k=1}$, $10\leq i \leq
12$). Interestingly, in the original stock network, stocks with a
large number of links to other stocks have a higher survivor ratio
than those with a smaller number of links. These finding are
confirmed in Fig. 1(d), which presents changes in the survivor ratio
with estimated returns and residual returns, respectively, as the
number of links with other stocks increase. In stock networks with
estimated returns, the survivor ratio of stocks with a large number
of links with other stocks have a higher value than those with a
smaller number of links. Meanwhile, in stock networks with residual
returns removing the properties of common economic factors, the
survivor ratio of stocks with a large number of links have a lower
value and zero value. That is, common economic factors have a major
and direct influence on stocks with a large number of links to other
stocks. These results suggest that common economic factors in the
stock market are the deterministic factors of the formation of the
stock network. Furthermore, stocks that act as a hub for clusters in
a network are affected more by common economic factors, while some
stocks with a small number of links, located on the outskirts of the
stock network, are affected less by common economic factors.

\section{Conclusions}

Stock markets have long been considered to be extremely complex
systems, which evolve through interactions between units. They have
also been known to form  homogeneous stock groups with a higher
correlation among different stocks according to common economic
factors that influence individual stocks. We investigated whether
common economic factors can be deterministic factors in the
formation of a stock network using the multi-factor model to reflect
the properties of common economic factors sufficiently, using the
daily prices of 400 individual stocks in the S\&P 500 index of the
American stock market from January 1993 to May 2005.

We have discovered empirically that the process of formation in a
stock network is significantly affected by common economic factors.
That is, the survivor ratio between original stock networks with
real returns and estimated stock networks with estimated returns had
a higher value as the number of common factors increased.
Furthermore, we found that stocks with a large number of links to
other stocks are affected more by common economic factors. That is,
the survivor ratio of stocks with a large number of links to other
stocks in the original stock network have a higher value than those
with a smaller number of links. Also, the highest consistency is in
stocks with the largest number of links to other stocks. These
results suggest that common economic factors in the stock market are
significantly deterministic factors in terms of formation principle
of a stock network. Additionally, stocks that act as a hub in a
network are affected by common economic factors, especially by the
market index, while stocks with a small number of links, located on
the outskirts of the stock network, are affected less by common
economic factors. Therefore, common economic factors in the stock
market are important deterministic factors of a stock network.

\begin{acknowledgements}
This work was supported by the Korea Research Foundation funded by
the Korean Government (MOEHRD) (KRF-2006-332-B00152), the MOST/KOSEF
to the National Core Research Center for Systems Bio-Dynamics
(R15-2004-033), the Korea Research Foundation (KRF-2005-042-B0075),
the Ministry of Science and Technology through the National Research
Laboratory Project, and the Ministry of Education and Human
Resources Development through the program BK21.
\end{acknowledgements}

\end{document}